\documentclass{iopart}

\usepackage{graphicx}
\usepackage{color}
\usepackage{float}
\usepackage{ulem}
\usepackage[english]{babel}
\usepackage[T1]{fontenc}

\newcommand{\csixty}{\mathrm{C}_{60}}
\newcommand{\wcmq}{\mathrm{W}/\mathrm{cm}^2}
\newcommand{\tpulse}{T_\mathrm{pulse}}

\newcommand{\omlas}{\omega_\mathrm{las}}
\newcommand{\epskin}{\epsilon_\mathrm{kin}}
\newcommand{\epol}{\mathbf{e}_\mathrm{pol}}
\newcommand{\ez}{\mathbf{e}_z}
\newcommand{\bohr}{\mathrm{a}_0}
\newcommand{\fwhm}{\mathrm{FWHM}}

\begin{document}


\title{On the dynamics of photo-electrons from C$_{60}$}

\author{C.-Z.~Gao$^{1,2}$, P.~Wopperer$^{1,2}$, P.~M.~Dinh$^{1,2}$, E.~Suraud$^{1,2}$, 
and P.-G.~Reinhard$^3$}
\address{$^1$
Universit\'e de Toulouse; UPS; Laboratoire de Physique Th\'{e}orique, IRSAMC; F-31062 
Toulouse Cedex, France}
\address{$^2$
CNRS; UMR5152; F-31062 Toulouse Cedex, France
}
\address{$^3$
Institut f\"ur Theoretische Physik, Universit\"at Erlangen, Staudtstra\ss e 7, D-91058 Erlangen, Germany}

\ead{dinh@irsamc.ups-tlse.fr}

\begin{abstract}
We explore photo-electron spectra (PES) and photo-electron
  angular distributions (PAD) of C$_{60}$ with time-dependent density
  functional theory (TDDFT) in real time.  To simulate
  experiments in gas phase, we consider isotropic ensembles of cluster
  orientations and perform orientation averaging of the TDDFT
    calculations.  First, we investigate ionization properties  of C$_{60}$ by
  one-photon processes in the range of VUV energies.  The PES map the
  energies of the occupied single-particle states, while the weights of
  the peaks in PES are given by the depletion of the corresponding
  level. The different influences can be disentangled by looking
  at PES from slightly different photon frequencies. PAD in
    the one-photon regime can be characterized by one parameter, the
  anisotropy. This single parameter unfolds worthwhile information
    when investigating the frequency and state dependences.  We
  also discuss the case of multi-photon ionization induced by
  strong infrared laser pulses in C$_{60}$. In agreement with measurements, we
  find that the PES show a regular comb of peaks separated by the
    photon energy. Our calculations reveal that this happens because
  only very few occupied states  of C$_{60}$ near the ionization threshold
  contribute to emission and that these few states happen to
    cooperate filling the same peaks. The PAD show a steady increase
  of anisotropy with increasing photon order. 
\end{abstract}

\submitto{\jpb}
\maketitle

\section{Introduction}

Early investigations of clusters concentrated on structure and optical
response \cite{deH93,Bra93,Kre93,Hab94a,Hab94b}.  Steady progress of
laser physics and experimental analysis has given access to more
detailed measurements as photo-electron spectroscopy (PES)
\cite{Gos83aB}, photo-electron angular distribution (PAD)
\cite{Coo69}, and in combination as PES/PAD, often represented
also as Velocity Map Imaging (VMI) \cite{Epp97}. These techniques have
been taken up intensely in cluster physics delivering a  richer
  information on the system than the mere ground
state single electron density of states \cite{Hui13}, see e.g. \cite{Fen10}. The more recent investigations can
address various dynamical regimes ranging from linear to nonlinear
domain by variations of the laser properties (i.e., intensity, duration, and frequency). 
%
The much celebrated fullerene
C$_{60}$, which could be regarded as one of archetypical clusters appearing between small molecules and bulk (solid) materials, has been at the core of numerous
investigations since several years. Studies on C$_{60}$ show the same development as for
clusters in general. They started out with considerations on
existence, structure and optical response
\cite{Kro85,deV92,Her92,Kro97}. First studies on PES came up also
rather early \cite{Lic91a,Ben91,Lie95}. More recently, a great
manifold of precision measurements of PES and/or PAD have been
performed, exploring several dynamical ranges, see
e.g. \cite{Hui13,Cam00,Han03,Kor05,MHL09,Kje2010,Joh12}.

The development of
theory is meanwhile parallel to the advance of measurements. Among the great variety of models, the best compromise between
expense and detailed description is achieved by Density Functional
Theory (DFT) for the electrons, coupled to the carbon ions by
pseudopotentials. This was already used for early investigations of
structure \cite{Tro92} and optical response \cite{Yab96}.  
Recently, anisotropy parameters (characterizing the 
PAD) have been extracted by using explicit ions and eigenfunctions of the scattering 
matrix in the framework of static DFT wave functions, see e.g.~\cite{Kor10,Tof10,Tof11}. Similar calculations
have  also been performed by employing Linearized Time-Dependent DFT (L-TDDFT), in which free plane waves for the outgoing wave functions are considered in the calculation of the
anisotropy parameter $\beta_2$, for various laser frequencies~\cite{Joh12}. However, free plane waves
might constitute an oversimplication which can significantly impact $\beta_2$, see discussion in 
Section~\ref{sec:wpad} and \cite{Wop13}. Other calculations of photo-ionization cross-sections based on L-TDDFT also 
exist but using a spherical hollow shell jellium background instead of an explicit ionic 
structure~\cite{MHL09,Mad08,McC08}.
The jellium approximation has also been used in~\cite{Ver12,Bol12,Ver13} to calculate cross-sections, but at the level of classical electrodynamics and hydrodynamics.
Unfortunately, the jellium approximation suffers from intrinsic limitations~:
It cannot reproduce the electronic shell closure
at $N_\mathrm{el}=240$ where it should be and it yields
a wrong sequence of single-particle (s.p.) levels.
As discussed in our previous works on Na clusters in the one-photon
regime, PES, PAD and anistropy parameters furthermore significantly differ
when explicit ions are used instead of a jellium background~\cite{Wop10a,Wop10b}.
This is particurlarly surprising as Na clusters possess a much better metallic character than the
partly covalent C$_{60}$~: the jellium model should therefore work better in Na clusters than in C$_{60}$.
We thus see as compulsory to use an explicit ionic structure of C$_{60}$ (without adjustable parameters)
for a quantitative description of PES and PAD, especially for a direct comparison with experimental data. 

The aim of this paper is thus to present in detail a model
providing a fully fledged microscopic theoretical
description of $\csixty$ using methods of
Time-Dependent DFT (TDDFT) in real time and real space, extensively applied and
tested previously in metal cluster dynamics \cite{Wop10a,Poh00,Poh04b,Wop12c}.
We will mainly address electronic emission from one-photon processes
induced by VUV pulses, but we will also have a look at strong infrared pulses
and subsequent multi-photon processes. 
Where possible, we shall perform comparison with available experiments.

The paper is outlined as follows: Section~\ref{sec:frame} briefly
introduces the theoretical framework. 
Section~\ref{sec:properties} discusses electronic and ionic structure
of C$_{60}$ and presents results on basic ground-state properties,
particularly those which are crucial for a proper description of
ionization processes, such  as Ionization Potential~(IP) and level
degeneracies.  
Section~\ref{sec:response} is devoted to the optical absorption
spectrum.
Section~\ref{sec:results} embraces all results on PES/PAD from
excitation by photon pulses in the monophoton regime. In this context, we also explain our
procedure for calculation of PAD and PES of an ensemble of randomly
oriented clusters.
Section~\ref{sec:multi} briefly explores the multiphoton regime.
Section~\ref{sec:conclusion} finally completes the paper with
a summary.

\section{Theoretical framework}\label{sec:frame}

\subsection{On the formalism}

The method used here, namely TDDFT in real time and real space coupled to 
a classical description of ions via pseudopotentials, is  standard  and 
has been presented in detail  in several papers and books, see  e.g. \cite{Cal00,PGR04}.
We shall thus only outline the specificities of the present investigation. 
We refer the reader to \cite{Cal00} for technical details.

The C$_{60}$ cluster is described in terms of 240 active valence
electrons. The coupling to the 60 carbon ions (one C nucleus with 2
core electrons each) is treated by non-local pseudopotentials of
Goedecker form~\cite{Goe96}.  The ions are kept frozen at the ground
state structure, which is legitimate for the short time scales
considered here.

Electron dynamics is described by TDDFT at the level of the Time-Dependent
Local-Density Approximation (TDLDA) propagated directly in the time
domain and using the $xc$-functional of \cite{PW92}.  The LDA is
augmented by a self-interaction correction \cite{PeZ81}. This is
crucial to achieve a relevant dynamical picture of emission properties.  In
practice, we use an average-density self-interaction correction \cite{Leg02}.  This formalism
is simple and yet reliable for a broad variety of molecules including
covalently bound systems \cite{Klu13a}.  To analyze observables from
electronic emission, we apply absorbing boundary conditions following
\cite{Cal00,PGR06}.

A word is in order about the limits of TDLDA in the dynamics of strong
photon excitations. A mean-field description cannot account for
dynamical correlations by electron-electron collisions. These convert
incoming energy into intrinsic heat which is released with large delay
by thermal electron emission. Thus, TDLDA does not account at all for thermal electron emission
  and overestimates direct electron emission. However, experiments collect all electrons emitted from
a cluster without discriminating arrival times, thus also including
thermal electrons. This has to be kept in mind in the
comparison with data later on. Thermal effects are so far mainly
described phenomenologically, see e.g. \cite{Han03}. A detailed
account of collisional thermalization is presently only possible in
semi-classical models of clusters dynamics \cite{Dom98b,Fen04}, which,
however, work so far only in metal clusters.  A
  quantum-mechanical modeling of thermalization has been proposed
  recently \cite{Rei14a} which is, however, still too elaborate to be
  used for a large system as C$_{60}$. One should finally mention that, on top of electronic thermal effects, the 
impact of ionic temperature should be added. Experimentally speaking, C$_{60}$ are produced at finite temperature, 
typically several hundreds of K, which implies sizeable temperature effects at the side of ions. 
This is in principle not a problem for TDDFT which can perfectly accommodate an ionic temperature 
in the course of the dynamics and ideally lead to a description in terms of an ensemble 
of TDDFT evolutions. Still, because C$_{60}$ is a large object, a proper description in terms of an 
ensemble of TDDFT computations remains extremely demanding in terms of computational cost. We shall 
thus recur to the zero temperature case all over the applications.

\subsection{Numerical aspects}

The numerical solution of the (time-dependent) Kohn-Sham equations for the
cluster electrons proceeds with standard techniques as described in
\cite{Cal00,PGR04}. The electronic wave functions and the spatial
fields are represented on a Cartesian grid in three-dimensional
coordinate space with grid spacing 0.71\,$\bohr$ and total box size of
$(112 \times 0.71\,\bohr)^3$. The absorbing boundary conditions are taken
spherical with the use of a radially symmetric mask function,
which is active only in the absorbing zone beyond
$|\mathbf{r}|>R_1=33$ a$_0$.  The mask switches gently from 1 at
$|\mathbf{r}|=R_1$ to 0 at the edges of the box.
The spatial derivatives are evaluated
via fast Fourier transformation. The ground-state configurations were
found by accelerated gradient iterations for the electronic
wave functions~\cite{Blu92}.  The ionic ground state configuration is
constructed from the known symmetries of C$_{60}$ \cite{Kro85,Dav91,Fri96},
while the ionic radius is optimized by minimizing the ground state
energy. Dynamical propagation   is done by the time-splitting method for
the electronic wave functions \cite{Fei82}.

\subsection{Excitation and observables}

Photon excitation of the cluster electrons is described by an
external dipole field $V_{\rm ext}$ which reads~:
\begin{equation}
V_\mathrm{ext} (t) =E_0 \, \mathbf{e}_\mathrm{z}\cdot\mathbf{r} \, 
\sin(\omlas t) \sin^2 \left(\pi t/\tpulse \right) \quad,
\label{eq:laser}
\end{equation}
and which is active in the time interval $0\leq t\leq\tpulse$.  The pulse
length $\tpulse$ relates to the full width at half maximum ($\fwhm$) in 
terms of laser intensity, $\fwhm\simeq\tpulse/3$.  We
will use short femtosecond (fs) pulses with moderate intensity and
propagate a while after the pulse is over, typically up to
$t=2\tpulse$, in order to collect most of the emitted electrons.

Three (experimentally measured) observables will be computed~:
spectral distribution of dipole strength, photo-electron spectra
(PES), and photo-electron angular distribution (PAD), the latter sometimes
combined as PES/PAD.  The dipole strength is obtained by excitation of
all occupied single particle wave functions with an instantaneous
boost $\varphi_\alpha(t=0) \mapsto
\exp(\mathrm{i}\mathbf{p}_\mathrm{boost}\!\cdot\!\mathbf{r})\varphi_\alpha(t=0)$
and subsequent spectral Fourier Transform (FT) analysis of the emerging dipole moment of the
electronic density \cite{Cal00,Cal97}.  The strength
$\mathbf{p}_\mathrm{boost}$ of the boost is chosen such that the
system remains safely in the linear regime with very small ionization (typically
10$^{-2}$). PES and PAD are investigated in connection with a laser pulse
 (\ref{eq:laser}) of finite length.
PAD are computed from accumulating the amount of absorbed electrons in
angular bins of the absorbing zone~\cite{Poh04b}.  PES are obtained by
storing the time evolution of the electronic wave functions at
selected measuring points close to the absorbing bounds and finally
Fourier transforming the collected signal~\cite{Poh00,Di12}.  Combined
PES/PAD are evaluated as PES, however, using a dense grid of measuring
points to obtain a sufficient angular resolution.
As clusters are usually observed in gas phase, we practically deal with
isotropic ensembles of cluster orientations. This requires orientation
averaging over observables collected from different orientations. This procedure requires a
specific methodology and dedicated calculations which were recently
explored in Na clusters in a series of papers \cite{Wop10a,Wop12c}.
The actual procedure used here will be outlined
in section \ref{sec:orientaver}.

To better analyze the details of the ionization dynamics,
we shall also complement the computation of the above observables by
the (asymptotic) depletion of each Kohn-Sham occupied level. Since we
use absorbing boundary conditions through a mask function, the norm of
each Kohn-Sham orbital $\varphi_i({\bf r},t)$ decreases with time.  We
thus define the depletion of each state as
$\sigma^{(i)}=1-\int\mathrm{d}\mathbf{r}|\varphi_i(\mathbf{r},t\rightarrow\infty)|^2$.
It can be shown that the $\sigma^{(i)}$'s are directly connected to
the PES and provide interesting details on the ionization mechanism
\cite{Vid10}.

\section{Ground-state properties}
\label{sec:properties}

\begin{figure}[htbp]
 \centering\hfill
 \raisebox{-0.5\height}{\includegraphics[width=0.3\linewidth]{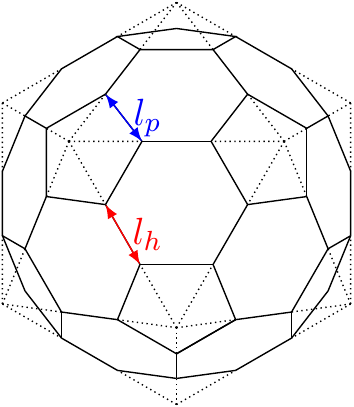}}\hfill
 \raisebox{-0.5\height}{\includegraphics[width=0.355\linewidth]{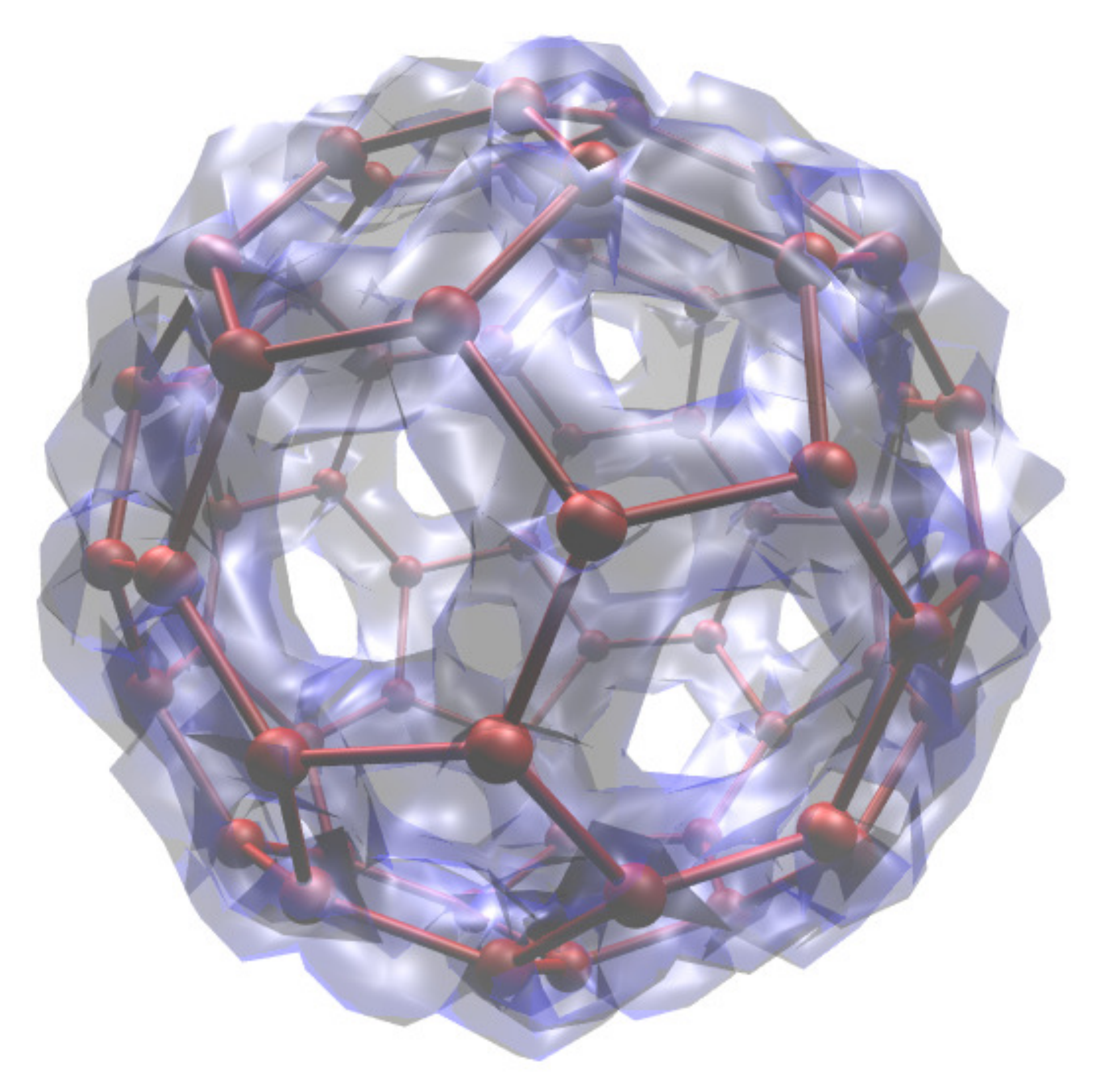}}\hfill
 \caption{Left: Ionic structure of
   C$_{60}$ with side lengths $l_h=2.759\,a_0$, $l_p=2.915\,a_0$, and
   radius $R_s=6.763\,a_0$.  Right:~Isosurface plot of the
   calculated electronic density.  }
 \label{fig:c60ionstruct}
\end{figure}

The C$_{60}$ cluster is distinguished by its highly symmetrical
structure, often denoted as a fullerene \cite{Kro85}.  The ionic
configuration is illustrated in the left panel of
Figure~\ref{fig:c60ionstruct}.
The structure can be described as a truncated icosahedron
\cite{Dav91,Hed91}. It consists of 20 hexagons and 12 pentagons. Each
pentagon is surrounded by hexagons. The 12 pentagons constitute 6
pairs of opposite faces, the 20 hexagons constitute 10 pairs of
opposite faces \cite{Fri96}. All atoms lie on a sphere of radius
$R_s$. The icosahedron (= 20 regular triangles) is truncated in such a
way that the resulting pentagons are regular with side length $l_p$
and the hexagons are irregular with two side lengths $l_p$ and $l_h$.
After structure optimization within our DFT calculation,
we get the values $l_h=2.759\,a_0$, $l_p=2.915\,a_0$, and
$R_s=6.763\,a_0$. They are respectively 3-6\% and 0.6\% larger than the typical
experimental values, that is, $l_h=2.648\,a_0$, $l_p=2.755\,a_0$, 
and $R_s=6.721\,a_0$~\cite{Hed91}.
They can be considered as sufficiently close to each other for the present purpose.

The right panel of Figure~\ref{fig:c60ionstruct} shows an isosurface
plot of the calculated electronic density in C$_{60}$.  This
density concentrates well around the ions. It also follows
predominantly the connecting lines, thus producing low-density region
at the center of the hexagons and pentagons, and leaving a large void
in the center of the cluster. Although we have these pronounced
low-density blobs, the electron channels are thoroughly connected,
which suffices to deliver the typical metallic behavior of free
electron flow throughout the whole cluster surface.

\begin{figure}[htbp]
 \centering
 \includegraphics[width=0.8\linewidth]{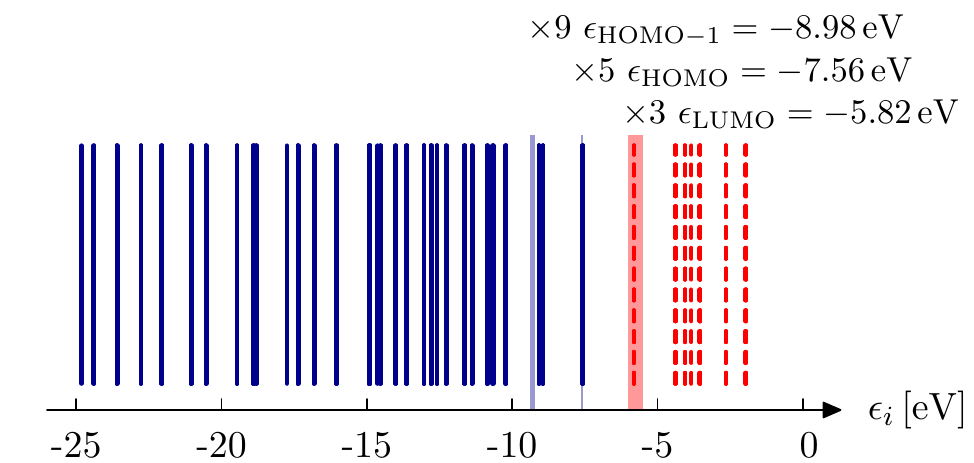}
 \caption{\label{fig:spspectrum} Calculated spectrum of single particle levels together
   with the degeneracy of HOMO, HOMO$-1$, and LUMO states. Experimental values
of the LUMO~\cite{Satt10}, HOMO~\cite{deV92,Zim90,Lic91b} and HOMO$-1$~\cite{Lie95,Satt10} 
are indicated as vertical shaded boxes.
   }
\end{figure}

Figure \ref{fig:spspectrum} shows the spectrum of single-particle
(s.p.) states. 
The occupied ones are indicated by full lines and the first few
unoccupied states by dashed lines.  Spin-up and spin-down states are
degenerated, thus leaving 120 occupied levels. The high symmetry of
C$_{60}$ leads to many degeneracies in the spectrum. We indicate in
the figure the degree of degeneracy (without spin-degeneracy) for the
most important states near the Fermi surface: the HOMO level is
five-fold, the HOMO$-1$ nine-fold, and the LUMO triply
degenerate. These degeneracies are also commonly obtained within
H\"uckel molecular orbital theory when taking into account the
icosahedral symmetry \cite{Had92}. The
HOMO-LUMO gap comes out at 1.74 eV in our calculations, in agreement
with experimental values mostly within the range from 1.6 to 2.1~eV~\cite{Satt10}. 
The energy difference from HOMO to HOMO$-1$ is of 1.42 eV,
while experimental values are 1.6--1.8 eV \cite{Lie95,Satt10}. This is
again a satisfying quality of reproduction in view of the complexity
of C$_{60}$. The computed IP is 7.56 eV, which nicely agrees with the
experimental value of 7.6 eV \cite{deV92,Zim90,Lic91b}. Note that a good
reproduction of the IP is crucial for a proper description of
photo-emission processes that we aim at here. 

\section{Optical response}
\label{sec:response}

\begin{figure}[htbp]
 \centering
 \includegraphics[width=0.8\linewidth]{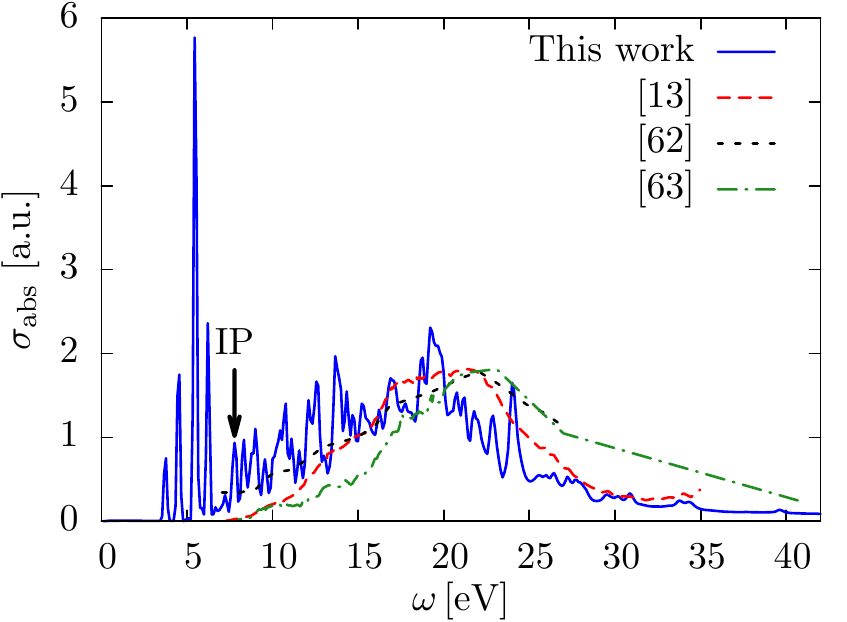}
 \caption{\label{fig:dip-resp-comp}
   Calculated dipole strength distribution $\sigma_\mathrm{abs}$ of C$_{60}$ (full line)
 for the mode in $z$-direction (vertical direction in the left panel of
   Figure~\ref{fig:c60ionstruct}), compared with experimental measurements~\cite{Her92,Kel92,Yoo92}. 
The calculated ionization potential is indicated by the arrow.}
\end{figure}

Figure \ref{fig:dip-resp-comp} shows the optical absorption (dipole)
strength.
Actually, this is the strength for excitation in $z$-direction
(symmetry axis = axis through the center of
a hexagon). The $x$- and $y$-modes are identical due to the high
symmetry of the C$_{60}$ cluster.  
Below the IP = 7.56 eV (left of the
vertical arrow), the dipole response exhibits several discrete
peaks. The lowest excitation energy of 4.1 eV is larger than the
HOMO-LUMO gap. In the lower energy regime, there is one peak at 6.0 eV
which dominates the other ones. It can be related to a collective
resonance of the $\pi$ electron system ($\pi$ plasmon). Its position
agrees well with other TDLDA calculations \cite{Yab99} and
measurements~\cite{Smi96,Yas96} (not shown on the figure to
    keep the very structured part below IP readable).
Above the IP, the response is dominated by
a broad Mie plasmon resonance around 19.2 eV. It is much fragmented
due to a coupling to the many one-particle-one-hole states in the
vicinity of the resonance, a mechanism which is called Landau
fragmentation, in analogy to Landau damping in a plasma
\cite{Lif88}. This resonance is very broad with a fragmentation width
(plus a sprinkle of continuum width) as large as 10 eV.
Generally, the lower modes (below 14 eV) correspond to excitations of
the highest occupied levels and have a $\pi - \pi^*$ character, while
transitions from the $\sigma$ subsystem also come into play at higher
energies. The Mie plasmon resonance at 19.2 eV can be ascribed to a
plasmon of all valence electrons ($\sigma+\pi$ plasmon).  
%
The experimental dipole strengths in Figure~\ref{fig:dip-resp-comp} show a
smooth, broad plasmon peak centered in the same region, 
while the calculated distribution
follows in the average the same trend but with larger
fluctuations. This is a consequence of a mean-field description at
fixed ionic configuration.  Dynamical correlations beyond mean field
lead to collisional broadening.  Moreover, clusters are usually
produced and measured at finite temperatures, e.g. at
300$^\circ$C~\cite{Kel92} or at 600$^\circ$C~\cite{Yoo92,Smi96}.  This
induces thermal fluctuations of the cluster shape which, in turn,
might also broaden the electronic excitations. Although thermal
broadening is less dramatic in the rather rigid C$_{60}$ cluster
than in the much softer metal clusters \cite{Mon95d}, both
broadening mechanisms together suffice to significantly smoothen the
strength distribution~\cite{Ell95}.
%

\section{Photo-emission in the one-photon regime}
\label{sec:results}

We now turn to situations where the free cluster is excited by photon
pulses as they are usually provided at synchrotron facilities.  In
this section, we consider linearly polarized pulses with high
frequencies ranging from 14\,eV to 34\,eV (which is far above the IP),
but with weak intensities of about $10^{10}\,\wcmq$.  In contrast to
synchrotrons, which deliver rather long pulses, we simulate the
dynamics with short femtosecond (fs) pulses of length in the order of
$\sim 50\,$fs for reasons of computational cost.  The photon
excitation leads to a weak total ionization between 0.001 and 0.1. The
dynamics thus safely stays in the perturbative regime with one-photon 
emission.

\subsection{Orientation averaging scheme}\label{sec:orientaver}

Experiments are performed in gas phase on an ensemble of randomly
oriented clusters. The isotropy of the ensemble simplifies the
PAD. For mere one-photon processes, it takes the form
\begin{equation}
  \frac{\textrm d\sigma}{\textrm d\cos\vartheta} =
  \frac{\sigma}{4\pi}\left[1+\beta_2 P_2(\cos\vartheta)\right]\:,
\label{eq:PADlin}
\end{equation}
where $P_2$ is the second Legendre polynomial, $\vartheta$ the
(a\-zi\-mu\-thal) emission angle measured with respect to the photon
polarization axis, $\sigma$ the emission yield, and $\beta_2$ the
anisotropy parameter. The PAD is symmetric under the following transformation
$\vartheta\leftrightarrow\pi-\vartheta$, and independent of the polar
angle $\varphi$. The anisotropy parameter $\beta_2$ can take values between $+2$
(pure $\cos^2$-shaped emission aligned with the photon polarization)
and $-1$ (pure $\sin^2$-shaped emission perpendicular to the photon
polarization).

\begin{figure}[htbp]
 \centering
 \includegraphics[width=0.5\linewidth]{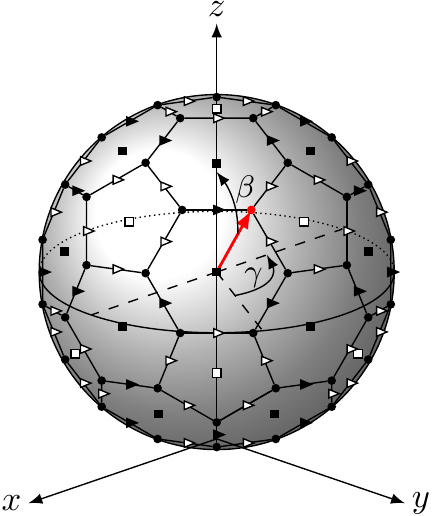}
 \caption{Illustration of the orientation averaging scheme. The Euler
   angles $\beta$ and $\gamma$ are shown for a sample cluster
   orientation (red arrow and red dot).  For C$_{60}$, 182 well
   chosen points reduce to 5 different orientations indicated by dots,
   filled and unfilled triangles and squares (only points in the foreground are displayed).}
 \label{fig:averaging}
\end{figure}

In theory, the situation is different. Here, we calculate the PAD of a
single cluster at fixed orientation.  Hence, comparison of theory with
experiment requires orientation averaging. An efficient scheme was
developed in \cite{Wop10a,Wop10b,Wop12a} which allows us to compute the
orientation-averaged PAD (OA-PAD). 
We use here a direct averaging which can be done by discretizing 
the integral of PAD over the sphere of orientations over 
a large selection of reference points. Each cluster orientation can be represented 
by a rotation of a given initial ionic configuration about the three Euler 
angles $\alpha$, $\beta$, and $\gamma$, as sketched in Figure~\ref{fig:averaging}. 
The final rotation (Euler angle 
$\alpha$) about the photon polarization axis $\epol=\ez$ does not require any
additional TDLDA runs and can be done a posteriori by averaging the
result over the polar angle $\varphi$ of the laboratory frame. This
leaves the Euler rotations $0\leq\gamma < 2\pi$ about the $z$-axis and
$0\leq \beta < \pi$ about the $y$-axis. Both angles can be drawn on
the unit sphere, see Figure~\ref{fig:averaging} for a sample rotation. 
The corresponding orientation and sampling point are indicated by the
red arrow and the red point, respectively. TDLDA calculations are then
performed for each of these configurations, the obtained PAD are
 added (with proper weight factors), and finally
averaged over the polar angle $\varphi$.

Although the direct averaging generally needs a lot of calculations
for a sufficiently converged result, the high symmetry of C$_{60}$ renders
direct orientation averaging competitive with only five orientations. 
Figure~\ref{fig:averaging} shows 
a set of 182 sampling points (only points in the foreground are
displayed) superimposed by the icosahedral structure of C$_{60}$.  For
this particular chosen setting, the sampling points can be divided
into 5 different groups of orientations which are connected by a
symmetry operation. Indeed, it reduces the number of needed
TDLDA calculations from 182 to 5 only.
 
Besides the calculation of the total OA-PAD of
Eq.~(\ref{eq:PADlin}), the above procedure can, of course, also be
used in order to calculate other orientation-averaged observables from
emission such as the total ionization  cross section  $\sigma$,
 s.p.~depletions
$\sigma^{(i)}$, angular distributions of s.p.~states
$\textrm d\sigma^{(i)}/\textrm d\!\cos \vartheta$,
photoelectron spectra  $\textrm d\sigma/\textrm d\epskin$, and
angle-resolved photoelectron spectra 
$\textrm d^2\sigma/(\textrm d\!\cos \vartheta\, \textrm d\epskin)$.
Results for each of these observables are presented in the
following.

\subsection{A first illustrative example}
\label{sec:example}

\begin{figure}[htbp]
 \centering
 \raisebox{-1.1\height}{\includegraphics[width=0.55\linewidth]{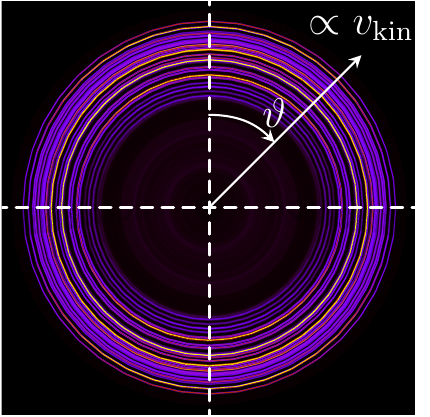}}
 \caption{Velocity map image obtained for one-photon
  excitation of C$_{60}$ with $\omlas=34\,$eV, $I=10^{10}\,\wcmq$, and
  $\tpulse=30\,$fs.
}
\label{fig:velomap25Ry-v6}
\end{figure}

Figure \ref{fig:velomap25Ry-v6} shows one typical example for
the full distribution
$\textrm d^2\sigma/(\textrm d\epskin\,\textrm d\!\cos\vartheta)$ obtained
in the mono-photon regime, visualized in a 3D projected map as the so-called
Velocity Map Image (VMI).
The map is drawn in polar coordinates where the angle $\vartheta$ stands for the
emission angle with respect to the laser polarization, 
and the radial coordinate for the
velocity $v_\mathrm{kin}=\sqrt{2\epskin/m}$ of the emitted electron. 
One observes a series of rings, each one standing for one
s.p.~state. Moreover, one realizes that each ring has its own intensity
(not necessarily isotropic) which means that different s.p.~states can have
different angular distributions \cite{Wop10a}.

\begin{figure}[htbp]
 \centering
 \includegraphics{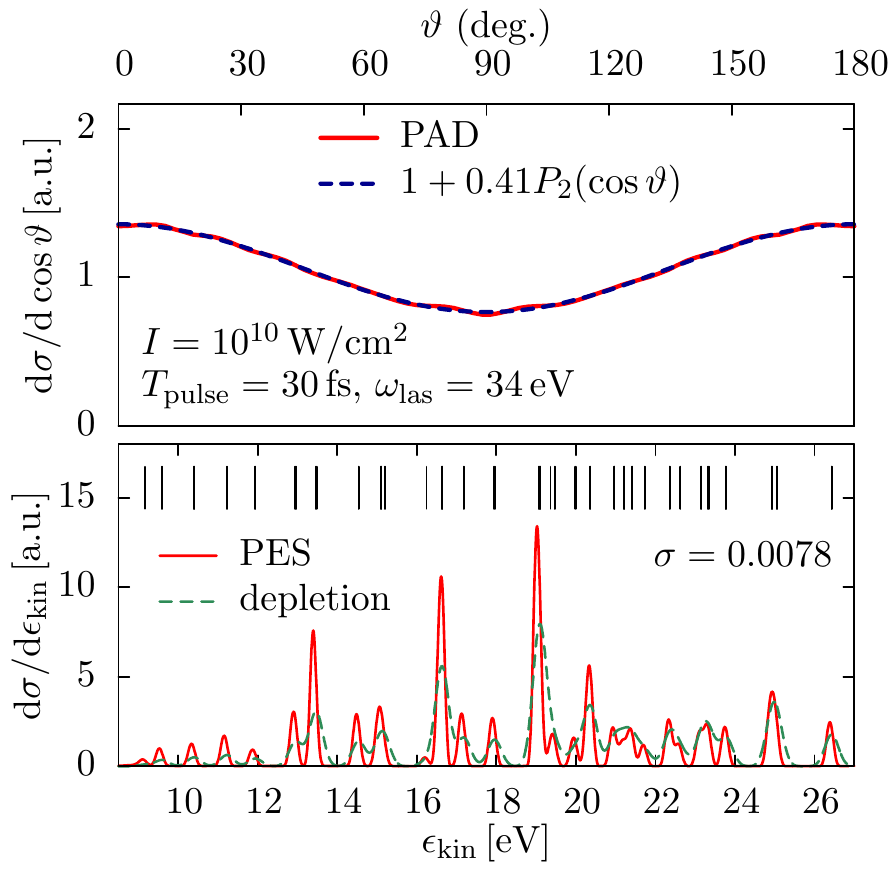}
 \caption{
  Orientation-averaged photoangular distribution (upper) and photoelectron spectrum 
(lower) obtained for one-photon
  excitation of C$_{60}$ with $\omlas=34\,$eV, $I=10^{10}\,\wcmq$, and
  $\tpulse=30\,$fs, with total ionization of $\sigma=0.0078$. 
The PAD is superimposed to the fit given in Eq.~(\ref{eq:PADlin}).  
The PES is superimposed to the single particle (s.p.) spectrum 
represented by Gaussians of constant width of 0.17~eV,
  weighted by the depletion of the s.p. states, and shifted by
  $\omlas$. The black  vertical lines in the lower panel indicate the
  (shifted) positions of the s.p.~states without weight.}
  \label{fig:pes-sync-theo2}
\end{figure}

Even if a VMI contains a priori richer information than PES or PAD
separately, it is hard to extract quantitative information from such an image. 
We thus discuss in the following the full distribution integrated over $\vartheta$,
that is the PES, and the distribution integrated over $\epsilon_{\rm kin}$, that is the PAD.
Figure \ref{fig:pes-sync-theo2} shows the PES 
and the OA-PAD for the same case as in Figure~\ref{fig:velomap25Ry-v6}.
The photon frequency is high enough to move all occupied
states directly into the continuum by 1-photon absorption, and the laser intensity is low 
enough to produce a very low total ionization of 0.0078, such that dynamics
stays in the linear regime.  The lower panel shows the PES.  When
comparing the PES peaks with the s.p.~levels as such (black vertical
lines at the top of the lower panel), we have the impression
that the PES somehow maps the s.p.~spectrum. However, the different
s.p. states obviously contribute with different weights to the
PES. This weight is related to the depletion of the state by the
photon pulse, which is a typically dynamical feature \cite{Vid10}.  We
thus weight the s.p.~states with their actual depletion (recorded
during the calculations) and give it a width corresponding to the
resolution set by the finite width of the photon pulse. The result
(denoted ``depletion'' in the figure) agrees perfectly with the
explicitly computed PES. It demonstrates that measurement of PES gives
indeed access to the spectra of the occupied states, however, weighted
with the depletion. The latter strongly depends on the photon
frequency \cite{Vid10}, see discussion below.  

As for the PAD in the top panel, the agreement of the computed OA-PAD with the second order polynomial
Eq.~(\ref{eq:PADlin}) is almost perfect, delivering an anisotropy parameter of $\beta_2=0.41$. 
By a closer inspection, one can spot
fluctuations about the smooth polynomial form. They stem from the
finite resolution of the sampling of the angular bins, but they remain
very small. This gives us confidence in our numerical scheme applied
to such a demanding system.

\subsection{Dependence on photon frequency}
\label{sec:wlas}

\subsubsection{Photoelectron and depletion spectra}

\begin{figure}[htbp]
 \begin{center}
 \includegraphics[width=0.8\linewidth]{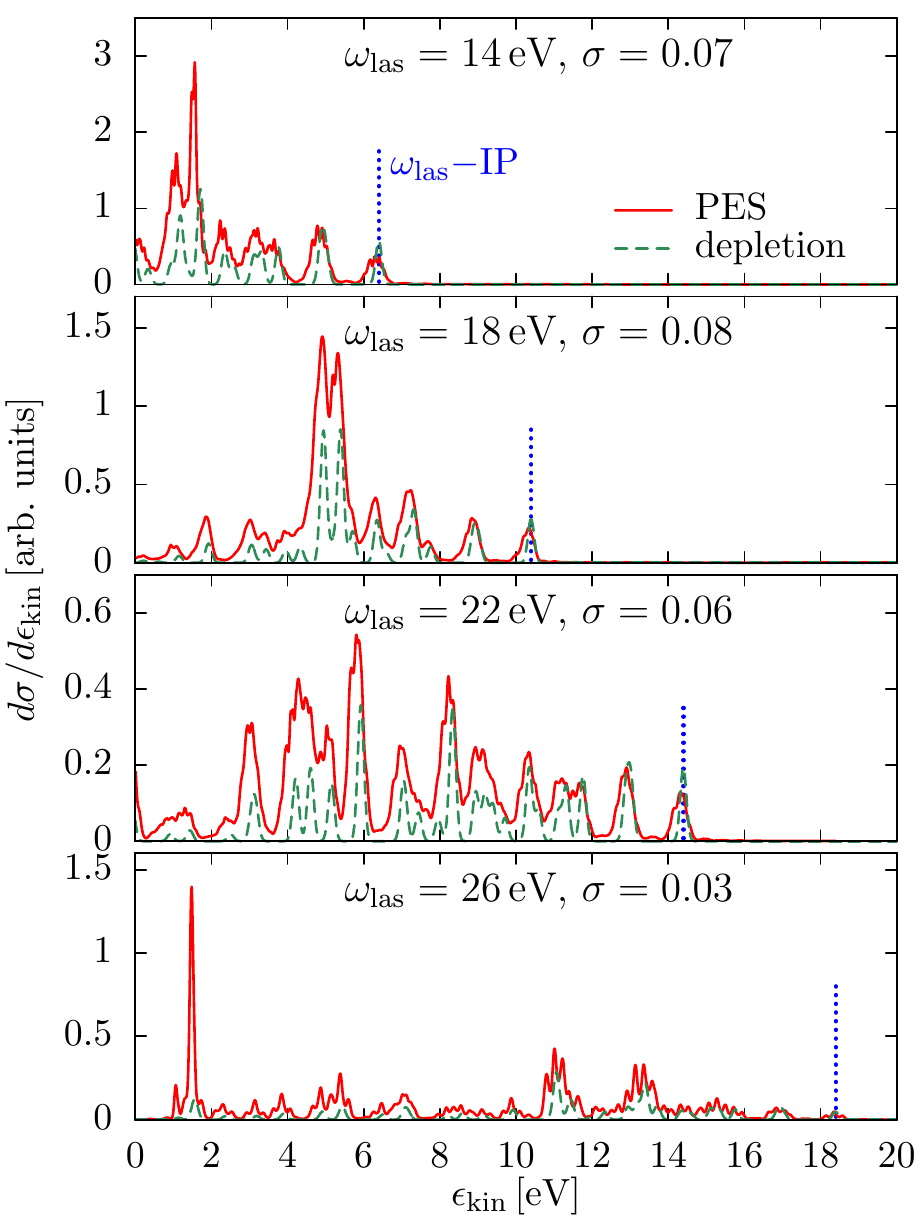}
 \caption{\label{fig:pes-deple} C$_{60}$ after irradiation by
a laser pulse of pulse length of 60~fs and intensity of $7.8\times 10^9\,\wcmq$, and
for different photon frequencies $\omega_{\rm las}$ and corresponding 
total ionization $\sigma$ as indicated~:
Photoelectron spectra (full lines) superimposed by single particle (s.p.) depletion spectra (dashes)
with a folding with Gaussians of constant width of 0.08~eV, as is done in the bottom panel
of Fig.~\ref{fig:pes-sync-theo2}. The dotted blue vertical lines indicate the threshold $\omega_{\rm las}-$IP
for one-photon ionization ($=$ ionization from the HOMO level).}
\end{center}
\end{figure}

We mentioned in connection with Figure~\ref{fig:pes-sync-theo2} that the
level depletion crucially depends on the photon frequency. 
In this section, we explore this effect.
To this end, we use photon pulses with different
frequencies, from 14 to 26~eV, and elsewise fixed parameters: pulse length of
60~fs and intensity of $7.8\times 10^9\,\wcmq$.
Figure \ref{fig:pes-deple} shows PES (full lines) and level depletions (dashes)
for four different photon frequencies
around the Mie plasmon peak $\omega_\mathrm{Mie}=19.2$ eV.
The spectra dramatically change with $\omega_{\rm las}$,
although the variation of frequency is not so large. 
Even if two-photon processes are in principle possible, they are much less likely than one-photon emission,
especially at this low laser intensity. 
Indeed, the spectra vanish above $\omega_{\rm las}-$IP in each panel of Figure~\ref{fig:pes-deple}. However, even if
the level sequences made accessible by one-photon processes remain the same
throughout all frequencies, the s.p. depletions significantly change  with
frequency. One measurement alone may by chance mask this or that
s.p. state.  Ideally, one should thus measure PES at a couple of
different and sufficiently high photon frequencies for a
reliable extraction of s.p.~spectra.  The sudden drop of signal
above the one-photon threshold $\omega_{\rm las}-$IP is particularly pronounced for weak
photon pulses. More intense pulses can more easily access deeper
levels.  Still, the strong dependence of depletion spectra 
remains also for other intensities (not shown here). 
At the side of the PES, the correlation with the depletion spectra
seems robust when the laser frequency varies. However
the PES signal does not trivially provide a one-to-one reproduction of
the s.p. density of state. It comes along overlayed with
dynamical features as, e.g., level depletion and pulse profile. 

\subsubsection{Photoangular distributions}
\label{sec:wpad}

\begin{figure}[htbp]
 \centering
 \includegraphics[width=0.8\linewidth]{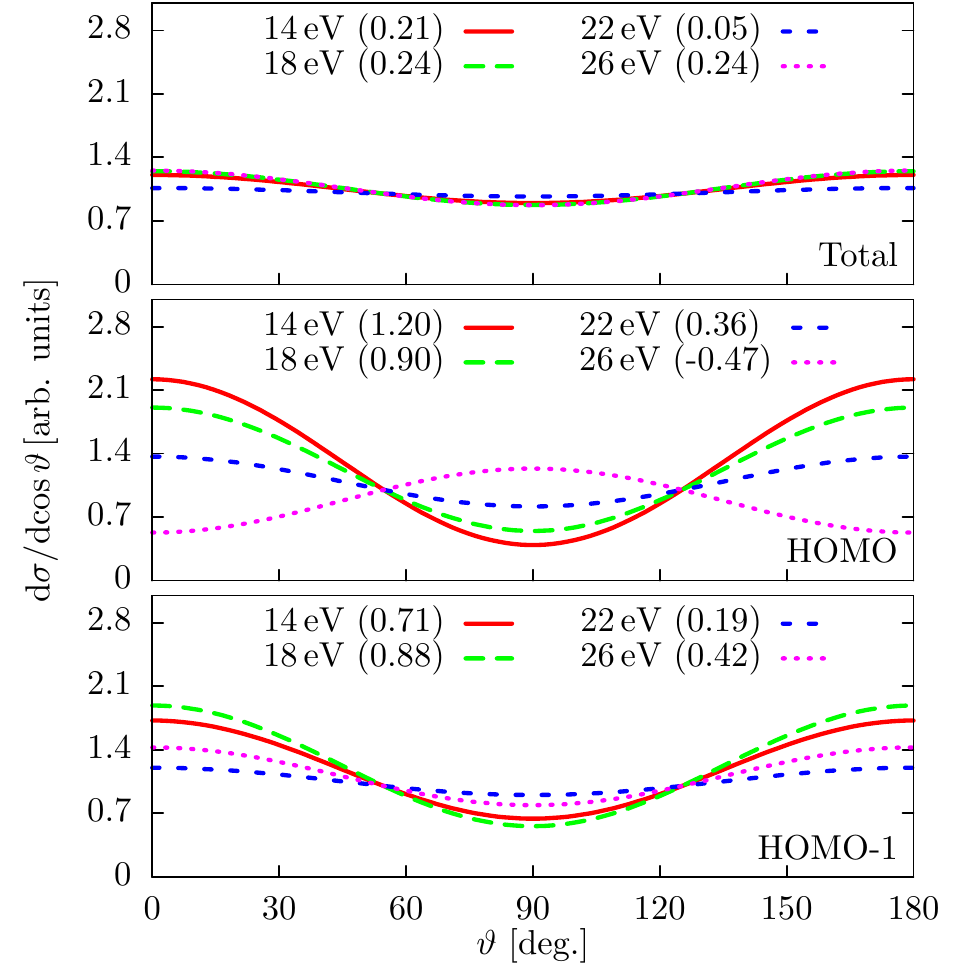}
 \caption{\label{fig:trend-omega4} PAD of all s.p. states (top), HOMO (middle), and 
 HOMO$-1$ (bottom) after irradiation by a laser with $I=7.8\times 10^9$~W/cm$^2$, 
$T_{\rm pulse}=60$~fs, and different photon frequencies as indicated. 
   Corresponding values for the anisotropy parameter $\beta_2$ are given in brackets.}
\end{figure}

We now turn to the frequency dependence of PAD.
Figure \ref{fig:trend-omega4} shows the PAD for different photon
frequencies, total PAD and selective for the group of HOMO and HOMO$-1$
levels. 
Note that the latter states are those which are experimentally
analyzed the best~\cite{Wop14}.  
In a given panel, each curve corresponds to one frequency and the number in 
parentheses is the $\beta_2$ resulting from a fit according to Eq.~(\ref{eq:PADlin}).
For total PAD (upper panel), the $\beta_2$ values do not change much with frequency and
are comparable to the test case at $\omega_{\rm las}=34$~eV in Section~\ref{sec:example}. 
This shows that a global PAD only provides a gross measure of the dynamics. Much more sensitivity is seen for the PAD for HOMO (middle panel) and \mbox{HOMO$-1$}
(lower panel). Both s.p.~anisotropies alone cover a larger range of values than the total
$\beta_2$. Even negative values (corresponding to emission preferably perpendicular to the
photon polarization) appear for the HOMO level at 26~eV. 
These larger variations of the s.p. anisotropies seem to be washed out when added up into the total anisotropy. State-resolved PAD, on the other hand, are much more sensitive to details of the system and dynamics.
They show sizeable variations from one level to the next and they depend significantly on
the laser frequency, a feature which we had already observed for the PES.

\begin{figure}[htbp]
 \centering
 \includegraphics[width=\linewidth]{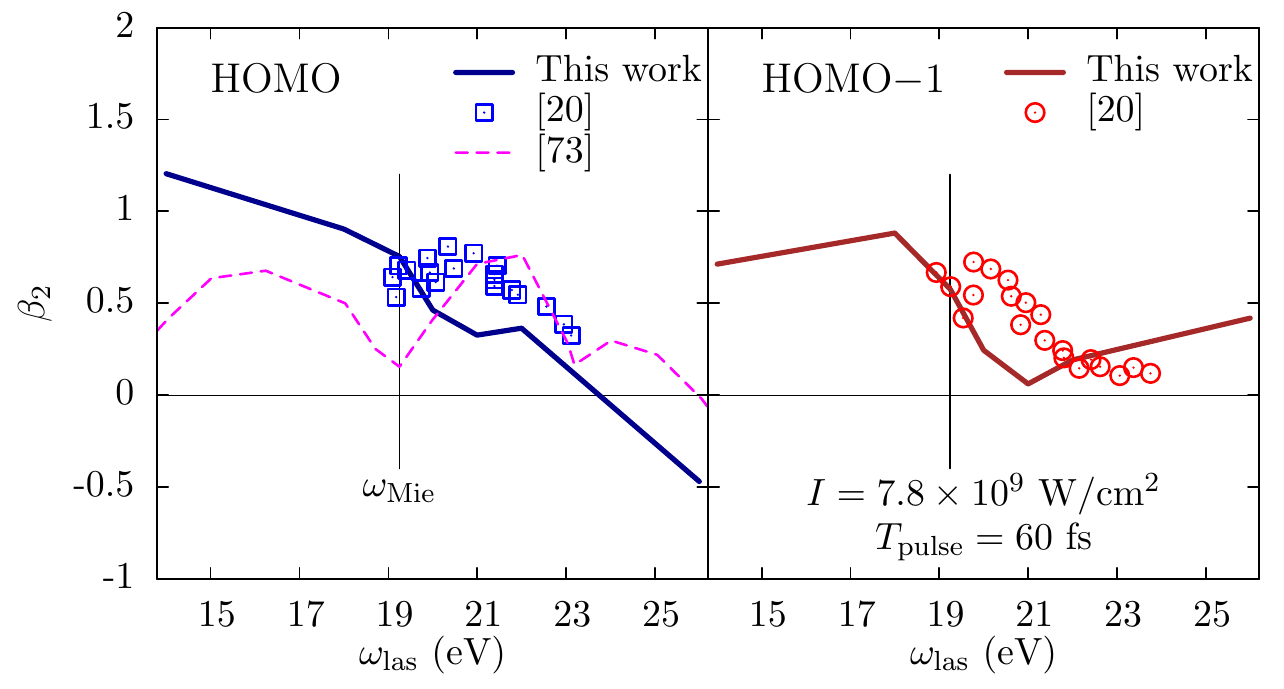}
 \caption{\label{fig:beta2_xuv} Anisotropy parameters
$\beta_2$ as a function of $\omlas$, 
from the HOMO (left panel), 
and from the HOMO$-1$ (right panel), with a comparison
of calculated values from this work (full lines) with experimental ones (symbols)~\cite{Kor05},
and with another theoretical work for the HOMO only (dashed line)~\cite{Gia01}.
The vertical line indicates the position of the calculated Mie plasmon frequency.}
\end{figure}

Figure~\ref{fig:trend-omega4} indicates that the PAD of HOMO and
 HOMO$-1$ vary significantly when going though the resonance energy,
 while the total PAD remains robust. Figure \ref{fig:beta2_xuv}
 analyzes that in terms of single particle $\beta_2$ on a denser mesh of frequencies
 $\omlas$ around the plasmon region $\omega_\mathrm{Mie}=19.2$~eV. 
It confirms these trends. Mind, however, that theoretical spectra in the
plasmon region, located above the IP, are fragmented into closely
packed sub-peaks, see Figure~\ref{fig:dip-resp-comp}. Each single
peak may produce resonant ionisation effects and significantly affect
the PAD. The situation is different in experiments where, in
particular, ionic temperature will lead to a smoothing of the optical
peaks into one broad and smooth structure around 20 eV. This may
soften quick variations of the PAD signal with laser frequency and
lead to a smoother behaviour altogether. 
Still, the comparison with previous experimental measurements~\cite{Kor05} is fairly good,
especially around $\omega_{\rm Mie}$.
The direct plotting of $\beta_2$ makes it also
clearly visible that the total $\beta_2$ is far from any average
between that of HOMO and that of HOMO$-1$.  There are indeed many more
states contributing substantially to the total $\beta_2$ as can read
off from the right panels of Figure~\ref{fig:pes-deple}.

To conclude this discussion, we also compare our obtained $\beta_2$ for the HOMO state with another calculation
from~\cite{Gia01}, see dashed curve in the left panel of Figure~\ref{fig:beta2_xuv}. 
These theoretical results are based on static calculations of matrix elements of the dipole operator
at the level of Hartree-Fock
molecular orbitals and the resolution of scattering equations by the single-center-expansion method.
The discrepancies with our calculations are large. However, they also significantly differ from the
experimental $\beta_2$. This demonstrates that $\beta_2$ is an extremely
sensitive observable at the side of the theory. And indeed, we have recently shown that ionization
cross-sections and anisotropy parameters strongly change when one uses as outgoing wave
functions either free plane waves, waves confined
in a square well or waves self-consistently calculated in a Kohn-Sham picture~\cite{Wop13}.

\section{An excursion into the multiphoton regime}
\label{sec:multi}

\begin{figure}[htbp]
 \centering
 \includegraphics[width=0.8\linewidth]{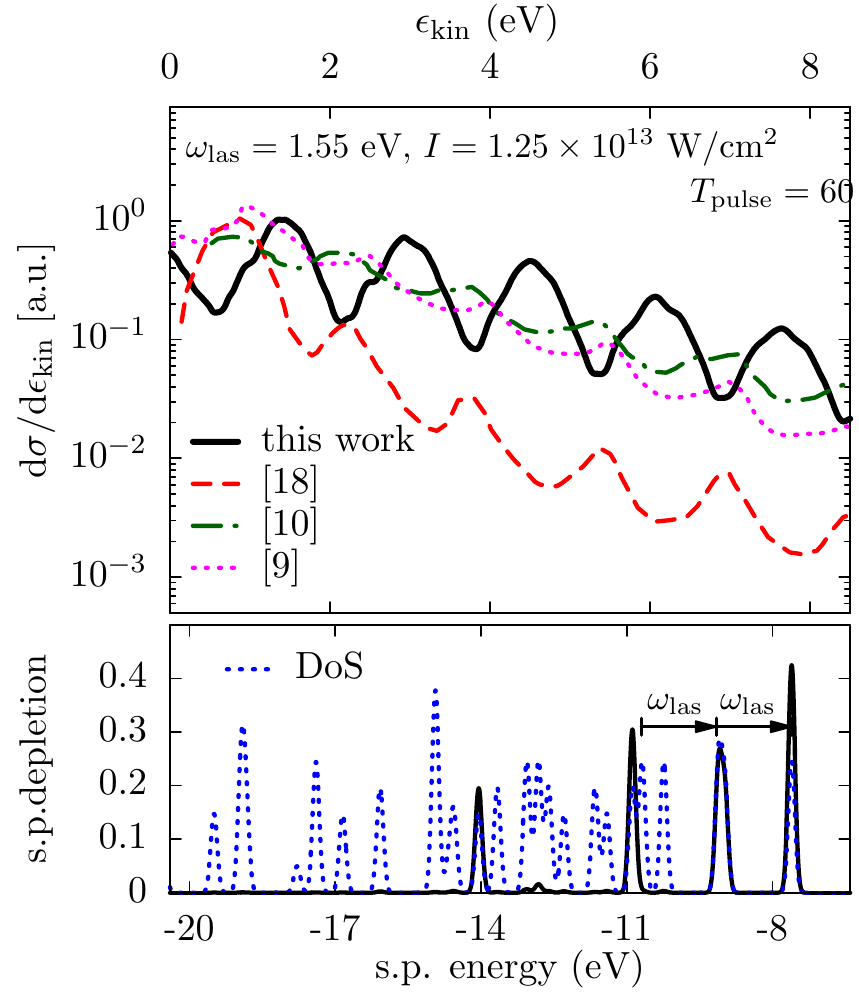}
 \caption{\label{fig:pes-deple_multi} Top~: calculated photoelectron
   spectrum (full line) with $I=1.25
\times 10^{13}$~w/cm$^2$, $\omlas=1.55$~eV, $T_{\rm pulse}=60$~fs
and compared with experimental measurements. Dashes~:$I=(8\pm2) 
\times 10^{13}$~w/cm$^2$, $\omlas=1.57$~eV, FWHM$=25$~fs~\cite{Cam00}~;
Dots and dashes~: $I=10^{13}$~w/cm$^2$, $\omlas=1.55$~eV, FWHM$=60$~fs~\cite{Fen10}~;
Dots~: $I= 2
\times 10^{13}$~w/cm$^2$, $\omlas=1.55$~eV, $T_{\rm pulse}=60$~fs~\cite{Hui13}. 
Bottom~: calculated depletion (solid line) of single
   particle states, compared with a (static) density-of-states (DoS)
   from all orbitals (dots).}
\end{figure}

Thus far, we have discussed one-photon processes using VUV pulses. 
At lower laser frequencies, multi-photon ionization (MPI) is
the way to electron emission. We consider here the example
of an infrared laser with $\omlas=1.55$ eV for which many experiments
have been done. The laser intensity has to be higher to achieve
an electron yield comparable with the monophoton case.
The top panel of Figure \ref{fig:pes-deple_multi} shows the calculated PES (full line)
compared with three different experiments of about similar laser parameters. 
The theoretical curve shows a clear peak structure with peak distance of
$\omlas$ and nearly exponential decrease of the envelope. The
exponential envelope is a known feature of MPI \cite{Poh04a}. The
surprise is here rather that we see such well developed peaks in
distances of 1.55 eV in spite of the fact that the s.p. spectrum is
spread densely over 17 eV $\gg\omlas$. The explanation is found in the
lower panel of Figure \ref{fig:pes-deple_multi}. The actual depletion
(full line) addresses only a few states near the HOMO while the rest
of the spectrum (see dotted line ``DoS'') remains untouched.
Moreover, these few states happen to line up fairly well with
$\omlas$. Thus the various multi-photon processes accumulate to the
same peak positions in the PES. For example, the first peak at 1.4 eV
is composed from a process with 6 photons from the HOMO, 7 photons
from HOMO$-1$, and 8 photons from the HOMO$-3$ group. 
The alignment of
frequencies is not perfect due to a slight mismatch of the HOMO$-3$.
This leads to
faint kinks in the theoretical peaks of the PES. Moreover, simple
photon counting would predict the first peak in the PES to appear at
1.7 eV. Actually, it is seen here at 1.4 eV.  This downshift, called
Coulomb shift, is an effect from ionization during emission. The
  total ionization is here about 0.06.  The thus higher charge
state enhances the Coulomb field of the cluster which, in turn, shifts
all s.p. energies to slightly deeper binding \cite{Poh00}.

The experimental PES are all recorded with slightly different laser
parameters. Nonetheless, experimental and theoretical results show all
in a similar fashion this typical regular sequence of peaks with
exponentially decreasing envelope. This indicates that the same MPI
mechanism is at work where only few levels near the HOMO contribute
whose energy difference matches approximately $\omlas$.  Going into
quantitative detail, we have to compare the key features of these PES:
peak positions, peak spacings, exponential slope, and amplitude of the
peaks. Peak spacings are the same for the four cases, but the three
other key features differ. This is not surprising as the
conditions are not exactly the same. But the differences demonstrate
how extremely sensitive PES depend on the dynamical conditions. 

For instance, the peak positions signal the ionization stage which crucially depends
on the IP, the actual laser intensity and the temporal pulse profile. However, neither
the laser intensity nor the pulse duration are precisely known experimentally, while these
two parameters strongly influence the final ionization stage and thus the Coulomb shift.
This is particularly visible when comparing the dotted~\cite{Wop14} and the
dashed-dotted~\cite{Fen10}
PES for which the laser intensities differ by 25~\%, and the
pulse duration maybe by a factor of 3. Therefore, even if the laser
parameters are very close in both experiments, the PES are almost out of phase.
The slope of the PES background also depends on the field strength (thus intensity)~: 
the higher the laser intensity, the smaller the slope~\cite{Poh04a}. 
However, if we take the experimental values of the laser intensity as such, we observe the reverse since
the largest intensity delivers the largest slope. Therefore this uncertainty on the  
laser intensities in an experiment once again hinders a quantitative comparison with the
theory. Note nevertheless a fair agreement of the theoretical slope with the experimental one 
from~\cite{Fen10} since in both
cases, one gets the same slope with a slight increase of the slope above $\epsilon_\mathrm{kin}\simeq 5$~eV.

Finally, the amplitude of the peaks is particularly involved, being composed from
broadening by Coulomb shift \cite{Poh00} and electron thermalization. It is
worth mentioning that in~\cite{Hui13}, a theoretical PES calculated within TDLDA but in the
jellium approximation is compared to an experimental PES using the same setup
of~\cite{Wop14} but at a slightly higher intensity ($I=2\times 10^{13}\,\wcmq$) and
measured at $90^\circ$ with respect to the laser polarization. The positions of the peaks
were satisfactory but the contrast of the MPI peaks was overestimated by the theoretical
calculations by more than two orders of magnitude. Here, we also get too large a peak
contrast in the theoretical curve but the discrepancy is less than a factor 10. This means
that explicit ions surely bring more dissipative effects at the side of electron-ion
collisions, and thus dramatically reduce the amplitude of the peak oscillations. Now, it
is still too large when compared with the experimental ones. Two complementing effects can
explain this difference. The missing of electron-electron collisions at TDLDA level
generally produces a larger amplitude. This is in particular an important effect at low
$\epsilon_{\rm kin}$. Moreover, we should emphasize that if the calculations are done at
0~K, this is far from being the case in the experiments, in which the temperature of the $\csixty$ jet ranges from 700~K to 850~K~\cite{Fen10,Cam00,Wop14}. 
And we know that a finite ionic temperature tends to blur high electron energy peaks and thus reduces signal to
background contrast~\cite{Wop14}. 

\begin{figure}[htbp]
 \centering
 \includegraphics[width=\linewidth]{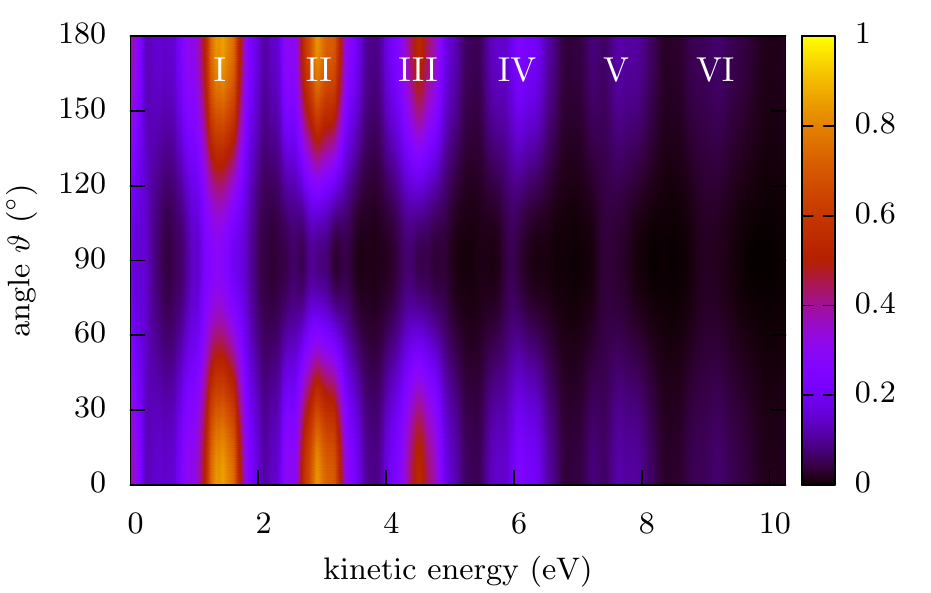}
 \caption{\label{fig:pespad_multi} 
Theoretical combined PES/PAD in the same multiphoton regime as in Figure~\ref{fig:pes-deple_multi},
that is $I=1.25\times 10^{13}$ W/cm$^2$, $\omega_{\rm las}=1.55$~eV, $T_{\rm pulse}=60$~fs.
Each peak is labeled by a roman letter (see text for details).
}
\end{figure}

We end up this discussion with Figure \ref{fig:pespad_multi} which shows a combined
PES/PAD, this time plotted in a rectangular plane of energy and
angle.
Remind that the first peak appearing in this plot corresponds to a 6-7-8 photon process. Each further
peak has one-photon-order higher. We see again the sequence of peaks
as in Figure~\ref{fig:pes-deple_multi}, which for completeness are identified with a roman number on the figure. 
New is the information on the angular distribution for each peak, which obviously depends on photon order. 
This is qualitatively clear form the figure but it can be made more quantitative by considering the $\beta_2$ values 
associated to each peak~: they are evaluated to be respectively $\beta_2=0.77$ for peak  I, $\beta_2=1.48$ for  II,
$\beta_2=1.50$ for III, $\beta_2=1.20$ for IV, $\beta_2=0.94$ for V and $\beta_2=0.92$ for VI. We thus roughly observe an increasingly forward emission as the number of involved photons increases.
This feature agrees well with available experimental data~\cite{Li13}.
Mind, however, that these $\beta_2$ are associated to rather large error bars of order of 0.5, and even 0.8
for the two highest energy peaks 
(V and VI) since they gather only little signal.  Thus the levelling off of $\beta_2$ 
 for point IV to VI should not be taken too serious.
Moreover, a thorough analysis of PAD in the MPI regime would require to
 look at higher order Legendre polynomials in the expansion
 (\ref{eq:PADlin}). This goes beyond the scope of the present paper.
Still, the important point is that the $\beta_2$ exhibits a sizeable dependence 
on the photon order, which once again shows how sensitive this quantity is.

\section{Conclusions}\label{sec:conclusion}

In this paper, we have investigated from a theoretical
  perspective ionization properties of C$_{60}$ irradiated by lasers
of various characteristics. To that end, we have used a fully
  microscopic description of electron dynamics based on time-dependent
  density-functional theory (TDDFT) coupled to the (frozen) ionic
  background via pseudopotentials. We have checked that this setup
delivers a good reproduction of ground state and low energy (optical
response) properties of C$_{60}$. We have then explored ionization
properties for the case of laser irradiations, considering the
  mono-photon regime as well as multi-photon processes. Thereby we
  concentrated particularly on energy- and angular-resolved emission
  cross-sections in terms of photo-electron spectra (PES) and
  photo-electron angular distributions (PAD). As experiments are
  usually done with C$_{60}$ in gas phase, we perform an orientiation
  averaging on all ionization properties to simulate the isotropic
  ensemble of orientations in the gas phase.

We have seen that both PES and PAD give access to structural
  properties of the cluster, however combined with dynamical features
  of the actual dynamical scenario. In the mono-photon regime, PES are expected
  to map the density of electronic states. In practice, this signal is
  folded with the depletion of each state and spectral profile of the
  laser. The case of C$_{60}$ is a priori particularly involved regarding its complex
s.p. spectrum. However, in every case studied in this paper, the PES peak positions and heights nicely correlate with the s.p. depletions. 
We also observe that increasing the laser frequency allows one to extract electrons
from deeper and deeper states but not in a monotonous way. The symmetry of the orbitals
certainly plays a role in the photoemission mechanism.
The structure of PAD becomes very simple in the mono-photon regime,
  reducing information to only one number, the anisotropy $\beta_2$.
  Nonetheless, PAD are found to provide worthwhile complementing
  information when considering their trends, e.g., as energy-resolved
  PAD or in dependence of systematically varied laser frequency. Here
  we find that the total anisotropy $\beta_2$ takes moderate values
  varying only slowly with frequency, while the $\beta_2$ for HOMO
 and HOMO$-1$ separately span a larger range of values and are
  highly sensitive to laser frequency, particularly around
  resonances. This is consistent with the fact that many states participate to the photoemission,
and thus to the total $\beta_2$, and not only the least bound states.

In the multi-photon regime, PES and PAD carry again combined
  information on cluster structure and dynamics of the excitation
  process, here with a stronger bias to dynamical influences.  A major
  difference to mono-photon processes is that emission is concentrated
  to a few shells near the Fermi surface. 
 As a consequence, PES are very structured with a sequence of
  peaks separated by the photon energy.  A particularly clean
  structure is found if the photon energy matches the energy distances
  between HOMO, HOMO$-1$, and HOMO$-3$ shells.  This result is in good
  agreement with measurements. Energy-resolved PAD nicely show that
  emission becomes increasingly forward focused with increasing number
  of photons involved.

The present survey was performed at zero temperature and ignored
  dynamical correlations from electron-electron collisions. Both
  effects tend to smooth spectral distributions, PES and PAD. They
  need to be included for a detailed reproduction of data. Work in
  that direction is in progress.

\section*{Acknowledgments}
This work was supported by the Institut Universitaire de
France. C.Z.G. receives financial support from China Scholarship
Council (CSC) (No. [2013]3009). This work was granted access to the
HPC resources of IDRIS under the allocation 2014--095115 made by GENCI
(Grand Equipement National de Calcul Intensif), of CalMiP (Calcul en
Midi-Pyr\'en\'ees) under the allocation P1238, and of RRZE (Regionales
Rechenzentrum Erlangen).


\begin{thebibliography}{10}
\expandafter\ifx\csname url\endcsname\relax
  \def\url#1{{\tt #1}}\fi
\expandafter\ifx\csname urlprefix\endcsname\relax\def\urlprefix{URL }\fi
\providecommand{\eprint}[2][]{\url{#2}}

\bibitem{deH93}
{de Heer} W~A 1993 {\em Rev. Mod. Phys.\/} {\bf 65} 611

\bibitem{Bra93}
Brack M 1993 {\em Rev. Mod. Phys.\/} {\bf 65} 677--732

\bibitem{Kre93}
Kreibig U and Vollmer M 1993 {\em Optical Properties of Metal Clusters\/}
  vol~25 (Springer Series in Materials Science)

\bibitem{Hab94a}
Haberland H (ed) 1994 {\em Clusters of Atoms and Molecules 1- Theory,
  Experiment, and Clusters of Atoms\/} vol~52 (Berlin: Springer Series in
  Chemical Physics)

\bibitem{Hab94b}
Haberland H (ed) 1994 {\em Clusters of Atoms and Molecules 2- Solvation and
  Chemistry of Free Clusters, and Embedded, Supported and Compressed
  Clusters\/} vol~56 (Berlin: Springer Series in Chemical Physics)

\bibitem{Gos83aB}
Gosh P 1983 {\em {Introduction to photoelectron spectroscopy}\/} (New York:
  Wiley)

\bibitem{Coo69}
Cooper J and Zare R~N 1969 Photoelectron angular distribution {\em Lectures in
  Theoretical Physics: Atom Collision Processes\/} vol~11 ed Geltman S,
  Mahanthappa K and Brittin W (Gordon and Breach, New York)

\bibitem{Epp97}
Eppink A~T~J~B and Parker D~H 1997 {\em Rev. Sci. Instrum.\/} {\bf 68}

\bibitem{Hui13}
Huismans Y, Cormier E, Cauchy C, Hervieux P~A, Gademann G, Gijsbertsen A,
  Ghafur O, Johnsson P, Logman P, Barillot T, Bordas C, L\'epine F and Vrakking
  M~J~J 2013 {\em Phys. Rev. A\/} {\bf 88}(1) 013201

\bibitem{Fen10}
Fennel T, Meiwes-Broer K~H, Tiggesb\"aumker J, Reinhard P~G, Dinh P~M and
  Suraud E 2010 {\em Rev. Mod. Phys.\/} {\bf 82} 1793

\bibitem{Kro85}
Kroto H~W, Heath J~R, O'Brien S~C, Curl R~F and Smalley R~E 1985 {\em Nature\/}
  {\bf 318} 162--163

\bibitem{deV92}
{de Vries} J, Steger H, Kamke B, Menzel C, Weisser B, Kamke W and Hertel I~V
  1992 {\em Chem. Phys. Lett.\/} {\bf 188} 159

\bibitem{Her92}
Hertel I~V, Steger H, de~Vries J, Weisser B, Menzel C, Kamke B and Kamke W 1992
  {\em Phys. Rev. Lett.\/} {\bf 68} 784

\bibitem{Kro97}
Kroto H 1997 {\em Rev. Mod. Phys.\/} {\bf 69} 703

\bibitem{Lic91a}
Lichtenberger D~L, Nebesny K~W and Ray C~D 1991 {\em Chem. Phys. Lett.\/} {\bf
  176}

\bibitem{Ben91}
Benning P~J, Poirier D~M, Troullier N, Martins J~L, Weaver J~H, Haufler R~E,
  Chibante L~P~F and Smalley R~E 1991 {\em Phys. Rev. B\/} {\bf 44}(4)
  1962--1965

\bibitem{Lie95}
Liebsch T, Plotzke O, Heiser F, Hergenhahn U, Hemmers O, Wehlitz R, Viefhaus J,
  Langer B, Whitfield S~B and Becker U 1995 {\em Phys. Rev. A\/} {\bf 52} 457

\bibitem{Cam00}
Campbell E~E~B, Hansen K, Hoffmann K, Korn G, Tchaplyguine M, Wittmann M and
  Hertel I~V 2000 {\em Phys. Rev. Lett.\/} {\bf 84} 2128

\bibitem{Han03}
Hansen K, Hoffmann K and Campbell E~E~B 2003 {\em J. Chem. Phys.\/} {\bf 119}
  2513

\bibitem{Kor05}
Korica S, Rolles D, Reink\"oster A, Langer B, Viefhaus J, Cvejanovi\'c S and
  Becker U 2005 {\em Phys. Rev. A\/} {\bf 71} 013203

\bibitem{MHL09}
Maurat E, Hervieux P~A and L\'epine F 2009 {\em J. Phys. B: At. Mol. Opt.
  Phys.\/} {\bf 42} 165105

\bibitem{Kje2010}
Kjellberg M, Johansson O, Jonsson F, Bulgakov A~V, Bordas C, Campbell E~E~B and
  Hansen K 2010 {\em Phys. Rev. A\/} {\bf 81} 023202

\bibitem{Joh12}
Johansson J~O, Henderson G~G, Remacle F and Campbell E~E~B 2012 {\em Phys. Rev.
  Lett.\/} {\bf 108} 173401

\bibitem{Tro92}
Troullier N and Martins J~L 1992 {\em Phys. Rev. B\/} {\bf 46} 1754

\bibitem{Yab96}
Yabana K and Bertsch G~F 1996 {\em Phys. Rev. B\/} {\bf 54}(7) 4484--4487

\bibitem{Kor10}
Korica S, Reink\"oster A, Braune M, Viefhaus J, Rolles D, Langer B, Fronzoni G,
  Toffoli D, Stener M, Decleva P, Al-Dossary O and Becker U 2010 {\em Surf.
  Sci.\/} {\bf 604} 1940--1944

\bibitem{Tof10}
Toffoli D and Decleva P 2010 {\em Phys. Rev. A\/} {\bf 81} 061201(R)

\bibitem{Tof11}
Toffoli D, Stener M, Fronzoni G and Decleva P 2011 {\em Chem. Phys. Lett.\/}
  {\bf 516} 154--157

\bibitem{Wop13}
Wopperer P, Reinhard P~G and Suraud E 2013 {\em Ann. der Physik\/} {\bf 525}
  309

\bibitem{Mad08}
Madjet M~E, Chakraborty H~S, Rost J~M and Manson S~T 2008 {\em J. Phys. B: At.
  Mol. Opt. Phys.\/} {\bf 41} 105101

\bibitem{McC08}
McCune M~A, Madjet M~E and Chakraborty H~S 2008 {\em J. Phys. B\/} {\bf 41}
  201003

\bibitem{Ver12}
Verkhovtsev A, Korol A and Solov'yov A 2012 {\em Eur. J. Phys. D\/} {\bf 66}
  253

\bibitem{Bol12}
Bolognesi P, Avaldi L, Ruocco A, Verkhovtsev A, Korol A and Solov'yov A 2012
  {\em Eur. J. Phys. D\/} {\bf 66} 254

\bibitem{Ver13}
Verkhovtsev A, Korol A and Solov'yov A 2013 {\em J. Phys.~: Conf. Ser.\/} {\bf
  438} 012011

\bibitem{Wop10a}
Wopperer P, Faber B, Dinh P~M, Reinhard P~G and Suraud E 2010 {\em Phys. Lett.
  A\/} {\bf 375} 39

\bibitem{Wop10b}
Wopperer P, Faber B, Dinh P~M, Reinhard P~G and Suraud E 2010 {\em Phys. Rev.
  A\/} {\bf 82} 063416

\bibitem{Poh00}
Pohl A, Reinhard P~G and Suraud E 2000 {\em Phys. Rev. Lett.\/} {\bf 84}

\bibitem{Poh04b}
Pohl A, Reinhard P~G and Suraud E 2004 {\em Phys. Rev. A\/} {\bf 70}

\bibitem{Wop12c}
Wopperer P, Reinhard P~G and Suraud E 2013 {\em Ann. Phys. (Berlin)\/} {\bf
  525} 309--321

\bibitem{Cal00}
Calvayrac F, Reinhard P~G, Suraud E and Ullrich C~A 2000 {\em Phys. Rep.\/}
  {\bf 337} 493

\bibitem{PGR04}
Reinhard P~G and Suraud E 2004 {\em Introduction to Cluster Dynamics\/}
  (Wiley-VCH Verlag, Weinheim)

\bibitem{Goe96}
Goedecker S, Teter M and Hutter J 1996 {\em Phys. Rev. B\/} {\bf 54} 1703

\bibitem{PW92}
Perdew J~P and Wang Y 1992 {\em Phys. Rev. B\/} {\bf 45}(23) 13244--13249

\bibitem{PeZ81}
Perdew J~P and Zunger A 1981 {\em Phys. Rev. B\/} {\bf 23} 5048

\bibitem{Leg02}
Legrand C, Suraud E and Reinhard P~G 2002 {\em J. Phys. B: At. Mol. Opt.
  Phys.\/} {\bf 35} 1115

\bibitem{Klu13a}
Kl\"upfel P, Dinh P~M, Reinhard P~G and Suraud E 2013 {\em Phys. Rev. A\/} {\bf
  88} 052501

\bibitem{PGR06}
Reinhard P~G, Stevenson P~D, Almehed D, Maruhn J~A and Strayer M~R 2006 {\em
  Phys. Rev. E\/} {\bf 73} 036709

\bibitem{Dom98b}
Domps A, Reinhard P~G and Suraud E 1998 {\em Phys. Rev. Lett.\/} {\bf 81} 5524

\bibitem{Fen04}
Fennel T, Bertsch G~F and Meiwes-Broer K~H 2004 {\em Eur. Phys. J. D\/} {\bf
  29} 367

\bibitem{Rei14a}
Reinhard P~G and Suraud E 2014 {\em subm. Phys. Rev. A\/} ArXiV:

\bibitem{Blu92}
Blum V, Lauritsch G, Maruhn J~A and Reinhard P~G 1992 {\em J. Comput. Phys.\/}
  {\bf 100} 364

\bibitem{Dav91}
David W~I~F, Ibberson R~M, Matthewman J~C, Prassides K, Dennis T~J~S, Hare J~P,
  Kroto H~W, Taylor R and Walton D~R~M 1991 {\em Nature\/} {\bf 353} 147

\bibitem{Fri96}
Fripertinger H 1997 {\em J. Chem. Inf. Comp. Sci.\/} {\bf 37} 535

\bibitem{Fei82}
Feit M~D, jr J~A~F and Steiger A 1982 {\em J. Comput. Phys.\/} {\bf 47} 412

\bibitem{Cal97}
Calvayrac F, Reinhard P~G and Suraud E 1997 {\em Ann. Phys. (N.Y.)\/} {\bf 255}
  125

\bibitem{Di12}
Dinh P~M, Romaniello P, Reinhard P~G and Suraud E 2013 {\em Phys. Rev. A\/}
  {\bf 87}(3) 032514

\bibitem{Vid10}
Vidal S, Wang Z~P, Dinh P~M, Reinhard P~G and Suraud E 2010 {\em J. Phys. B:
  At. Mol. Opt. Phys.\/} {\bf 43} 165102

\bibitem{Hed91}
Hedberg K, Hedberg L, Bethune D~S, Brown C~A, Dorn H~C, Johnson R~D and
  de~Vries M 1991 {\em Science\/} {\bf 254}

\bibitem{Satt10}
Sattler K 2010 {\em Handbook of Nanophysics: Clusters and Fullerenes\/}
  Handbook of Nanophysics (CRC Press)

\bibitem{Zim90}
Zimmerman J~A, Eyler J~R, Bach S~B~H and McElvany W 1991 {\em J. Chem. Phys.\/}
  {\bf 94} 3556

\bibitem{Lic91b}
Lichtenberger D~L, Jatcko M~E, Nebesny K~W, Ray C~D, Huffman D~R and Lamb L~D
  1991 {\em Mater. Res. Soc. Symp. Proc.\/} {\bf 206}

\bibitem{Had92}
Haddon R~C 1992 {\em Acc. Chem. Res.\/} {\bf 25} 127--133

\bibitem{Kel92}
Keller J~W and Coplan M~A 1992 {\em Chem. Phys. Lett.\/} {\bf 193} 89

\bibitem{Yoo92}
Yoo R~K, Ruscic B and Berkowitz J 1992 {\em J. Chem. Phys.\/} {\bf 96} 911

\bibitem{Yab99}
Yabana K and Bertsch G~F 1999 {\em Intern. J. Quant. Chem.\/} {\bf 75} 55

\bibitem{Smi96}
Smith A~L 1996 {\em J. Phys. B: At. Mol. Opt. Phys.\/} {\bf 29} 4975

\bibitem{Yas96}
Yasumatsu H, Kondow T, Kitagawa H, Tabayashi K and Shobatake K 1996 {\em J.
  Chem. Phys.\/} {\bf 104} 899--902

\bibitem{Lif88}
Lifschitz E~M and Pitajewski L~P 1981 {\em {Physical Kinetics}\/} ({\em Course
  of Theoretical Physics\/} vol~X) (Oxford: Butterworth-Heinemann)

\bibitem{Mon95d}
Montag B and Reinhard P~G 1995 {\em Phys. Rev. B\/} {\bf 51} 14686

\bibitem{Ell95}
Ellert C, Schmidt M, Schmitt C, Reiners T and Haberland H 1995 {\em Phys. Rev.
  Lett.\/} {\bf 75}(9) 1731--1734

\bibitem{Wop12a}
Wopperer P, Dinh P~M, Suraud E and Reinhard P~G 2012 {\em Phys. Rev. A\/} {\bf
  85}(1) 015402

\bibitem{Wop14}
Wopperer P, Gao C~Z, Barillot T, Cauchy C, Marciniak A, Despr\'e V, Loriot V,
  Celep G, Bordas C, L\'epine F, Suraud E and Reinhard P~G 2014  Submitted

\bibitem{Gia01}
Gianturco F~A and Lucchese R~R {\em Phys. Rev. A\/} {\bf 64} 032706

\bibitem{Poh04a}
Pohl A, Reinhard P~G and Suraud E 2004 {\em J. Phys. B: At. Mol. Opt. Phys.\/}
  {\bf 37} 3301

\bibitem{Li13}
Li H 2013 {\em Study on molecular photoionization in femtosecond laser field\/}
  Master's thesis Kansas State University Manhattan, U.S.A.
  \urlprefix\url{https://krex.k-state.edu/dspace/handle/2097/15913}

\end{thebibliography}
\providecommand{\newblock}{}

\end{document}